\documentclass[sigconf]{acmart}

\AtBeginDocument{}
\setcopyright{none}
\copyrightyear{2026}
\acmYear{2026}
\acmDOI{}
\acmConference[]{Workshop on Computing within Limits, June 23-25, 2026}{}{}
\acmISBN{}

\begin{document}
\title{Urban Limits as Design Constraints: Identifying Suitable Locations for Distributed, Photovoltaic-Powered Servers}

\author{Justin Chikhaoui}
\email{justin.chikhaoui@lirmm.fr}
\affiliation{
  \institution{LIRMM, Univ. Montpellier - CNRS}
  \city{Montpellier}
  \country{France}
}

\author{Thomas Leduc}
\email{thomas.leduc@crenau.archi.fr}
\affiliation{
  \institution{Nantes Université, ENSA Nantes, École Centrale Nantes, CNRS, AAU-CRENAU, UMR 1563}
  \city{Nantes}
  \country{France}
}

\author{Daniel Siret}
\email{daniel.siret@crenau.archi.fr}
\affiliation{
  \institution{Nantes Université, ENSA Nantes, École Centrale Nantes, CNRS, AAU-CRENAU, UMR 1563}
  \city{Nantes}
  \country{France}
}

\author{Abdoulaye Gamati\'e}
\email{abdoulaye.gamatie@lirmm.fr}
\affiliation{
  \institution{LIRMM, Univ. Montpellier - CNRS}
  \city{Montpellier}
  \country{France}
}

\renewcommand{\shortauthors}{Chikhaoui et al.}

\begin{abstract}
Urban territories face growing tensions between increasing digital demand, limited resources, and socially constrained built environments. Although distributed computing paradigms such as edge and fog computing are widely presented as solutions for reducing latency and energy dissipation, the scientific literature largely overlooks where such infrastructures can be physically and socially deployed within cities, and typically neglects urban constraints, environmental impacts, and equity considerations.
	
This paper proposes a methodology for identifying suitable urban locations for deploying distributed servers under structural, environmental, and social limits. Relying exclusively on existing infrastructures and anthropised surfaces, it combines legal frameworks, ongoing urban projects, citizen consultations, and scientific literature to construct a place-based glossary of viable site typologies, evaluated through energy, spatial, and qualitative criteria. Applied to the French city of Montpellier, our results illustrate how urban constraints and local resources shape the feasibility of decentralised, solar-powered digital infrastructures, and highlight the value of territorialised approaches for rethinking digital services within urban limits.
\end{abstract}


\received{}
\received[]{}
\received[]{}

\maketitle

\section{Introduction}

Urban territories constitute a fundamentally complex framework, characterised by 
a dense and heterogeneous built environment, a plurality of actors, and a living 
space shaped by sometimes contradictory requirements \cite{shtob2025planning}.

Within this space, the \textit{production of the city} consists of responding to 
collective needs — education, health, culture, mobility, economic activities — 
while navigating material, social, and political constraints 
\cite{boeing2018measuring}. Increasingly, the response to these needs is mediated 
by digital services: sensors, applications, platforms, control systems, and 
decision-support tools. This digitisation dynamic sustains the idea that 
technology could respond ever more precisely to human needs 
\cite{matei2024artificial}, promising an enhanced understanding of urban systems 
and their actors, e.g., through land-use strategies that integrate a 
holistic view of the territory.

Recent decades have, however, seen the rise of a technosolutionist vision 
embodied by initiatives such as \textit{smart cities} \cite{wiig2015ibm}, which promote 
the massive extension of sensors and digital services as a privileged approach 
to solving urban and societal problems. This vision is not neutral: it entails 
significant environmental and social impacts \cite{williams2011environmental, 
yamin2019information}. Among the most widely deployed digital 
infrastructures, data centers require substantial material inputs during both construction 
and operation. The extraction, transformation, and transportation of these 
materials exert growing pressure on natural reserves and the territories where 
they are located \cite{zhu2020monitoring}. On the social dimension, the widespread 
deployment of digital devices reshapes daily practices, social relations, and 
conditions of access to urban services, fostering dependency on technical 
infrastructures and reinforcing the vulnerability of populations with limited 
digital access or skills \cite{raihan2025dimensions}.

These observations raise a central question: \textit{how can digital needs — not 
all of which are necessarily desirable — be met without exceeding the 
environmental and social limits of urban territories, limits that existing 
material infrastructures already displace and exacerbate?}

\paragraph{\bf Our vision}

Several strategies address the tension between meeting digital needs and managing 
their associated impacts \cite{perea2023digital, selwyn2024digital}: sobriety, 
degrowth, the abandonment of certain services, or the redefinition of needs, 
among others. Here, we focus on an approach that leverages specific 
characteristics of the urban environment. We identify at least two properties 
that enable such an objective: (1)~the \textit{judicious reuse} of the already 
urbanised and artificialised built environment; and (2)~the exploitation of 
urban solar resources. By ``judicious reuse'', we refer to a deployment model in 
which digital demand and the computational resources that serve it are 
co-located, while remaining within the bounds imposed by the integration of 
infrastructure into the built urban environment (notably its institutional, 
architectural, and social constraints).

Two further elements are considered in order to meet digital needs without 
deteriorating other dimensions of the territory. The first concerns the spatial 
limits of cities: our approach favours locally grounded responses to local 
needs, anchoring our work within the specific challenges of a fine-grained 
territorial scale through a bottom-up strategy \cite{theis2025local}. The second 
concerns the structural specificities of urban territories — building ownership, 
infrastructure functions, social acceptability, and so on — which call for a 
deployment approach that is sensitive to the living and working conditions of 
the territory, ensuring that computing units are integrated with due regard for local actors.In practice, this translates into the deployment of computing units powered by photovoltaic energy on existing infrastructures and buildings. Reusing existing buildings prevents further land artificialisation and the associated environmental impacts, while solar energy minimizes dependence on fossil fuels.This strategy may nonetheless generate unintended consequences 
\cite{ngata2025cloud}.The first is the potential proliferation of digital 
infrastructures and services. 
The second is the displacement of the primary social functions of the selected sites: urban furniture and public facilities serve as spaces of rest and collective appropriation that may be compromised by the technical requirements of hosting computing units, particularly in dense areas where public space is already limited. technological expansion effect, we propose deploying computing units only in 
situations of infrastructure renewal, thereby avoiding the creation of new 
infrastructures that do not respond to a specific need — a situation that could otherwise stimulate digital demand and reinforce the associated environmental and social impacts \cite{alcott2005jevons, luccioni2025efficiency}.  Furthermore, the intermittency of solar energy restricts continuous operation and increases maintenance needs, thereby reducing incentives for intensive exploitation and encouraging more restrained patterns of use. Together, these constraints lead us to select urban deployment locations with careful attention to the existing uses of infrastructures and buildings.

\paragraph{\bf Contribution and outline}

This paper makes a twofold contribution. The first is a methodology aligned with the vision presented above, based on the identification and characterisation of suitable locations in urban environments so as to better account for the limits and dynamics of urban spaces. The second is the application of this methodology through a case study focused on the city of Montpellier, in southern France.

The paper is organised as follows. We first review the relevant literature 
(Section~\ref{sec:related}), then detail the proposed methodology  
(Section~\ref{sec:methodo}), before presenting a case study that allows us to identify a typology of urban locations suited to the digital needs specific to Montpellier based on our methodology (Section~\ref{sec:case-study}). Finally, we give concluding remarks and discuss some future 
research directions and potential applications 
(Section~\ref{sec:conclu}).
\section{Placement of computing units: a problem mainly framed as 
algorithmic optimization}
\label{sec:related}

The scientific literature offers a substantial body of work seeking to address 
the tension between digital demand and its territorial impacts 
\cite{goel2024overview, finnveden2025assessing, lu2024digital, 
papadonikolaki2022digital} through the deployment of distributed architectures. 
In particular, \textit{edge} and \textit{fog computing} are presented as 
paradigms that bring computing capabilities closer to points of use, thereby 
reducing latency, network congestion, and the energetic distance between data 
production and processing \cite{Yousefpour2019}.

\paragraph{\bf Distribution of computing units: a strategy aimed at promoting sustainability}

Distributed infrastructures, such as edge computing systems, able to address the bandwidth and latency issues of centralized infrastructures, in particular by reducing congestion on long-haul networks such as submarine fibre-optic cables. They are often described as inherently more sustainable than the centralized ones, but this claim is contested. Several studies show that sustainability outcomes depend less on the spatial proximity of computing resources than on how infrastructure deployment, capacity provisioning, and the surrounding energy system are configured \cite{arroba2024sustainable, bharany2022systematic, kautish2025advancing, tocze2022dark}. The environmental benefits of distribution are therefore not intrinsic; they depend on these systemic choices.

The literature points to two main levers for improving the sustainability of edge and fog computing \cite{Yousefpour2019}: powering distributed infrastructures with renewable sources \cite{da2022optimization}, and co-locating computing units with energy production sites to limit transmission losses \cite{queiroz2012energy}. 
The distributed nature of edge and fog computing also makes it possible to exploit territorial heterogeneity, such as locally available renewables or favourable climatic conditions, and to rethink infrastructure design beyond network-centric criteria. Passive cooling through outdoor immersion of edge nodes has been evaluated \cite{guo2021advanced}, and other work proposes reusing the waste heat of distributed units to supply urban thermal demand \cite{dowds2021utilising}.

\paragraph{\bf Distribution of computing units: a strategy aimed at promoting sustainable development}

Distributed computing was initially conceived to address the bandwidth and latency limitations of centralized infrastructures, notably by reducing congestion in communication networks such as submarine fiber-optic systems.

Although distributed paradigms are often portrayed as inherently more sustainable than centralized computing, this assumption remains controversial. Several studies indicate that sustainability outcomes depend less on the spatial proximity of computing resources than on the interplay between infrastructure deployment, capacity provisioning, and the characteristics of the associated energy systems \cite{arroba2024sustainable, bharany2022systematic, kautish2025advancing, tocze2022dark}.

In this sense, the environmental benefits of distribution cannot be considered intrinsic, but rather contingent upon broader systemic configurations. In parallel, the literature identifies several strategies that could enhance the sustainability potential of edge and fog computing. As emphasized in the literature \cite{Yousefpour2019}, energy-efficient approaches are essential to mitigate the environmental impact of these distributed paradigms. Two main directions can be identified: on the one hand, the integration of renewable energy sources to power edge and fog infrastructures \cite{da2022optimization}, and on the other hand, the co-location of computing units with energy production sites, which may reduce transmission losses and improve overall energy efficiency \cite{queiroz2012energy}.

More broadly, the distributed nature of edge and fog computing offers the opportunity to leverage territorial heterogeneity, for instance by exploiting locally available renewable resources or favorable environmental conditions. This perspective enables a rethinking of infrastructure design beyond purely network-centric considerations. For example, passive cooling strategies based on the outdoor immersion of edge nodes have been explored \cite{guo2021advanced}, while other works propose reusing the waste heat generated by distributed computing units to meet urban thermal demands \cite{dowds2021utilising}.

\paragraph{\bf Geographical distribution of computing units: a central yet 
under-explored issue}

Surveys on distributed infrastructures consistently show that the literature 
has concentrated on architectures, resource management, security, and 
orchestration \cite{Mouradian2017, Abbas2017, Hu2017}, leaving a significant 
blind spot: the geographical positioning of computing units. Yousefpour et 
al.~\cite{Yousefpour2019} explicitly note that, despite its importance for 
quality of service and infrastructure sustainability, this topic remains 
under-studied due to the absence of standardised frameworks and mature 
protocols.

The works that do address geographical placement generally model it as a 
multi-objective optimization problem, integrating criteria such as latency, 
load balancing, cost, or energy consumption \cite{Cao_2021_Large-Scale, 
liu2022}. These approaches, however, share recurring limitations. They 
typically rely on pre-defined locations — base stations, roadside units, or 
access points — which restrict the range of sites considered 
\cite{asghari2024}. They also tend to adopt incomplete territorial models, 
which may yield theoretically optimal placements that are urbanistically 
infeasible \cite{Benamer_2021}. Digital demand is moreover often estimated 
indirectly, either through generic network communication flows 
\cite{wang2019delay} or population-based territorial models \cite{asghari2024}. 
Taken together, these tendencies reveal a persistent dissociation between 
computational optimization and territorial embeddedness — a dissociation 
that stands in tension with the necessity, recalled in the introduction, of 
articulating digital services with the material and social realities of the 
urban environment.

\paragraph{\bf Integrating the urban territory into the placement problem}

A number of recent studies have nonetheless begun to engage more seriously 
with the urban territory. Several highlight that existing infrastructures can 
constitute deployment opportunities for computing units \cite{laha2020, 
aral2021, madamori2021latency, gamatie2023}, offering advantages such as 
structural or energy accessibility, proximity to usage contexts, pre-existing 
electrical supplies or communication devices, and, in the case of rooftops, 
favourable solar exposure.

Beyond static infrastructure, other contributions incorporate spatial and 
temporal dynamics \cite{chen2021preference, shao2022, chang2021}. Population 
densities, variations in digital demand, and daily mobility patterns all affect 
where computing capacity is most relevant, and accounting for these dynamics 
helps avoid rigid deployments ill-suited to territories characterised by 
heterogeneous human flows and uneven digital practices.

A further strand of work underlines the importance of urban constraints — 
physical obstacles, architectural properties, and failure risks — which can 
affect signal propagation and network robustness \cite{shahin2021, shahin2023}. 
These studies make clear that the reliability of a distributed computing 
network is not solely an algorithmic property but also a material and spatial 
one. Finally, certain contributions, such as \cite{gamatie2023}, explicitly 
seek to articulate deployment with urban renewable energy, modelling the solar 
potential of rooftops to size locally powered computing nodes and demonstrating 
that a convergence between digital infrastructure and local energy resources is 
achievable.

\paragraph{\bf Limitations of existing work}

Despite these contributions, research on the urban placement of computing units remains fragmented across disciplines and approaches. Each study tends to isolate a single territorial 
dimension without offering a framework that encompasses the multiple material, 
social, and energetic layers conditioning deployment feasibility. No existing 
work aims to construct a broad, systematic, and characterised inventory of 
urban locations capable of hosting computing units.

Furthermore, the reliance on indirectly estimated digital demand introduces 
biases with potentially significant consequences. Oversizing decentralized 
infrastructures may encourage increased digital usage and triggering a 
technological pull effect. Conversely, undersizing them may fail to meet actual urban 
needs and may weaken essential territorial functions. Both outcomes can deepen inequalities among inhabitants, as some would see their needs 
adequately addressed while others remain marginalised from equitable access 
to digital services.

More broadly, the literature reveals three persistent gaps: the geographical 
placement of distributed computing units remains under-explored outside 
information and communication sciences; the specific characteristics of urban 
territories represent major yet still under-exploited levers; and existing 
approaches account for only a fraction of the dimensions required to design 
infrastructure that is truly situated, sustainable, and compatible with urban 
limits. 

Our work positions itself precisely within this gap, by simultaneously 
accounting for potential urban locations, their structural and energetic 
properties, and the nature of the digital needs that the infrastructure must 
address.

Beyond the technical literature on distributed computing, the LIMITS research community has already begun to address many of the questions raised in this work, particularly regarding how digital systems can be designed under environmental and social constraints. Prior work in computing within limits emphasizes the need to reconsider the growth-oriented paradigm of digital infrastructures, advocating instead for approaches grounded in sufficiency, resource awareness, and the acknowledgment of material limits \cite{tomlinson2013collapse, silberman2015information, nardi2018computing}. In this perspective, notions such as digital sobriety, rebound effects, and the redefinition of needs have been explored as central challenges for sustainable computing \cite{bates2018intangible}. Moreover, LIMITS research stresses the importance of situated and place-based approaches, where infrastructures are designed in relation to local contexts, resources, and communities rather than as abstract, globally optimized systems \cite{houston2022richness}. Our work builds on these conceptual foundations by operationalizing them in the context of urban digital infrastructure siting, proposing a methodology that explicitly integrates spatial, social, and energy constraints into the identification of deployment locations. In doing so, it contributes to bridging the gap between the critical, conceptual perspective developed within LIMITS and the more operational questions of infrastructure design and placement.

The LIMITS community has already addressed several of the questions this work raises, particularly how digital systems can be designed under environmental and social constraints. A prior work argues for moving away from the growth-oriented paradigm of digital infrastructures and toward approaches grounded in sufficiency, resource awareness, and material limits \cite{tomlinson2013collapse, silberman2015information, nardi2018computing}. Within this stream, digital sobriety, rebound effects, and the redefinition of needs have been examined as central challenges for sustainable computing \cite{bates2018intangible}. LIMITS research also calls for situated, place-based approaches, in which infrastructures are designed w.r.t. local contexts, resources, and communities rather than as abstract globally optimized systems \cite{houston2022richness}.

We build on these foundations and operationalize them for urban digital infrastructure siting, proposing a methodology that integrates spatial, social, and energy constraints into the identification of deployment locations. The contribution thus connects the critical perspective developed within LIMITS to the operational questions of infrastructure design and placement.

\section{Methodology for computing unit deployment}
\label{sec:methodo}

Our methodology aims to identify suitable locations for deploying distributed 
computing units in urban environments, motivated by the absence of a rigorous 
framework in the scientific literature for determining which sites are 
appropriate for such infrastructures. It guides location selection toward 
buildings and surfaces that respect the structural and environmental limits of 
the urban territory through two explicit constraints. The first is a 
\textit{material} constraint: reducing the environmental footprint of digital 
technologies by relying strictly on existing buildings and infrastructures. The 
second is a \textit{social} constraint: avoiding conflicts of use by selecting 
only locations whose technical or multipurpose function can accommodate a 
digital service without disrupting their primary role.

\begin{figure}[htbp]
    \centering
    \includegraphics[width=0.33\textwidth, trim={0 0 0 0},clip]
        {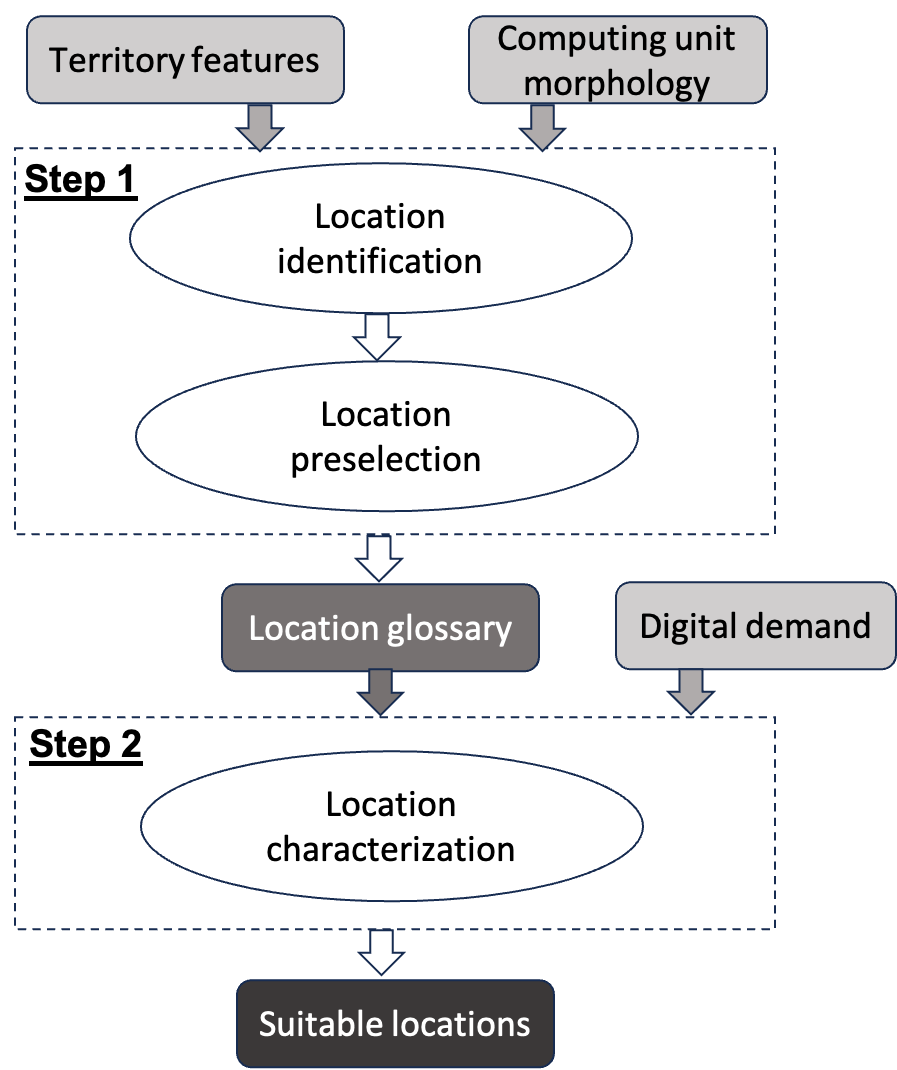}
    \caption{Methodology for identifying relevant locations for the deployment 
    of distributed server systems within the structural limits of the urban 
    territory.}
    \label{fig:Methodologie}
\end{figure}

The methodology unfolds in two steps (Figure~\ref{fig:Methodologie}). The first 
identifies candidate locations that satisfy urban acceptability constraints, 
producing a \textit{glossary} of viable site typologies. The second 
characterises these locations in order to select those best suited to digital, 
energy, and social needs. These steps must be understood within their 
territorial context: grounding the search in a specific territory makes it more 
realistic by incorporating local specificities. Legislation governing rooftop 
usage, for instance, differs between different countries. Similarly, 
the physical morphology of computing units conditions which locations are 
feasible: a small IoT-type device may be installed at a roadside traffic signal, 
whereas a server equipped with memory, storage, and a GPU requires a larger 
dedicated space.

The glossary produced by the first step constitutes a foundational element of 
our approach, directly addressing a gap in the literature on location types 
suitable for distributed server deployment. It takes the form of a dataset of 
locations characterised by their typology and accompanied, at minimum, by their 
geographical coordinates.

The second step takes as input this glossary together with a digital demand 
defined both spatially and quantitatively. Proper qualification of this demand 
is essential to ensure proportionality: an excessive deployment of servers risks 
inducing a rebound in digital consumption \cite{alcott2005jevons}, whereas 
insufficient deployment would fail to meet actual needs.

Through this methodology, we do not seek to optimise the deployment of 
distributed servers according to classical metrics — latency, performance, or 
load balancing. Rather, we aim to assess the actual capacity of the urban 
territory to host a form of digital technology that is more environmentally 
sober and socially acceptable.

\subsection{Identification of suitable locations}

The first step takes as input a specific territory features and a computing unit  morphology (e.g., servers) to 
be deployed; several morphologies may also be considered simultaneously.

\paragraph{\bf Identification}

Because the search for locations is bounded by both a territory and a server 
morphology, we focus on sites that may satisfy these two constraints. The 
methodology therefore draws on typologies of locations already in use within 
the territory for comparable deployments, identified through four sources: 
law and regulation, existing projects, citizen consultation reports, and 
scientific literature.

\begin{figure}[htbp]
    \centering
    \includegraphics[width=0.38\textwidth, trim={0 8cm 0 8cm},clip]
        {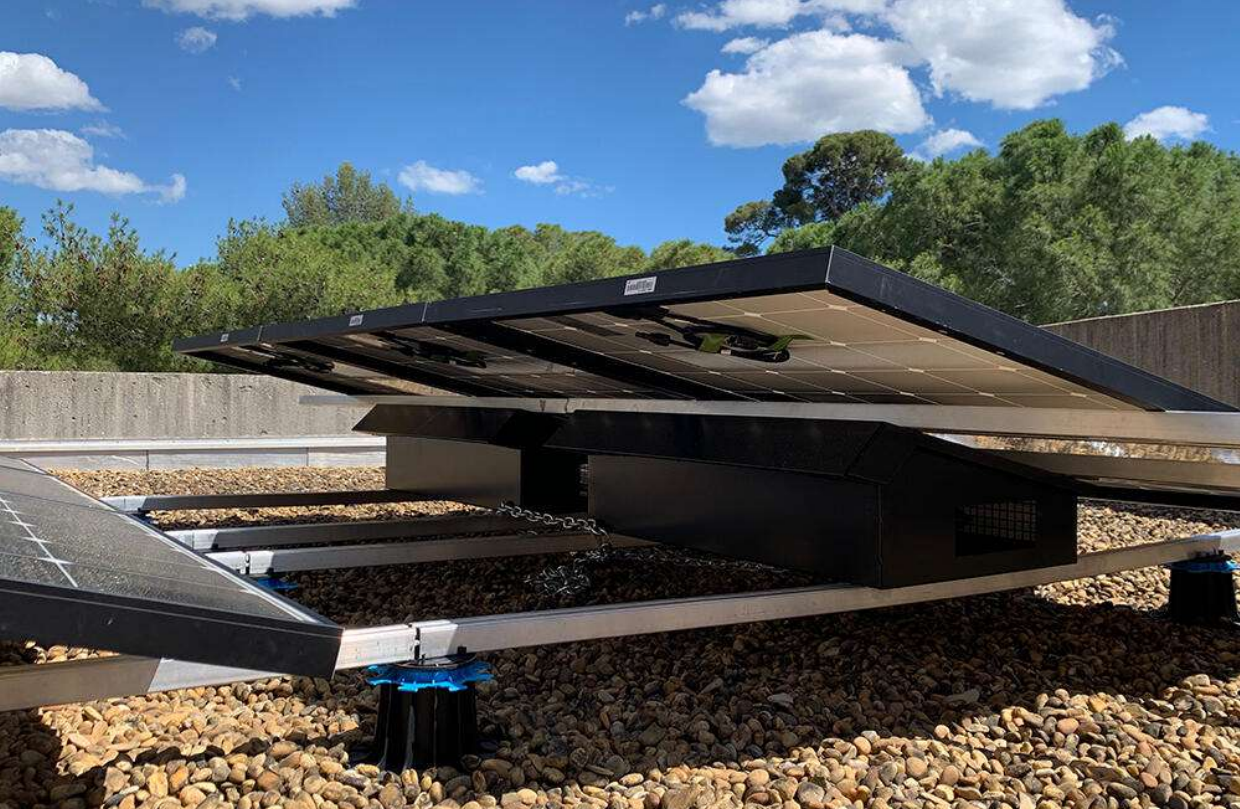}
    \caption{Genesis prototype deployed at the University of Montpellier}
    \label{fig:genesis_panel}
\end{figure}

Given that the deployment of distributed servers is rarely discussed explicitly 
in these sources, we adopt the following working hypothesis: \textit{servers may be 
installed in the same locations as solar panels, such that a panel may be accompanied by a computing unit without requiring 
additional space, as illustrated by Genesis systems 
\cite{sassatelli:lirmm-02145436, DBLP:journals/tsusc/SilvaGSPR23}.} 
Genesis (short for "\textit{Generation of Energy-Synergistic Information Systems}") (Figure \ref{fig:genesis_panel}) exemplifies a green computing paradigm. It is an electronic device that integrates at least one solar panel with energy transfer logic and can optionally include batteries and compute servers with corresponding data transfer capabilities. Its solar panels enable local energy harvesting, while the Genesis computing systems could also serve as a cloud-based Platform-as-a-Service (PaaS) infrastructure.

The Genesis paradigm addresses the resource-intensity computing services by aligning compute with the conditions of its host environment rather than expanding dedicated infrastructure. It has three properties that support this alignment. First, placing computing units next to photovoltaic production reduces transmission losses and lets capacity track the local energy profile. Second, outdoor deployment with a suitable enclosure provides natural ventilation, which removes most of the active cooling components and the energy they consume. Third, mounting units on existing urban infrastructure rather than building dedicated facilities avoids further urban sprawl, a direct driver of biodiversity loss.

\textit{Law and regulation} inform the obligations, recommendations, and 
constraints applying to any intervention in urban development 
\cite{bennett2016does}. Obligations may identify areas constrained for 
deployment; recommendations reveal locations where deployment would be 
supported by territorial governance; and constraints indicate sites where 
deployment would be prohibited or difficult. For example, Articles L111-19-1 
and R111-25-17 of the French urban planning code identify parking areas as 
locations where renewable-energy devices are required 
\cite{code_urbanisme_L111_19_1, code_urbanisme_R111_25_17}.

\textit{Urban projects} conducted by local authorities, transport operators, 
companies, and territorial organisations provide concrete grounding for the 
reflection on digital materiality and the need to rely on resources already 
present within a territory \cite{tsui2020role}. Rather than adding dedicated 
infrastructures, urban experiments show that cities already contain numerous 
locations capable of hosting energy devices and, by extension, distributed 
computing units \cite{desthieux2018solar}. They confirm that the relevance of 
a solar-first deployment depends on identifying existing, stabilised, 
low-conflict surfaces. Among such examples, the Wattway project demonstrates 
the feasibility of integrating photovoltaic panels directly into the pavement 
of cycling lanes \cite{wattway_colas}.

\textit{Citizen consultation} reports, although rarely accessible, provide 
insight into how inhabitants appropriate and reshape the territory to match 
their expectations. They therefore represent a source of locations that have 
been discussed and validated by those most directly concerned, often accompanied 
by motivations that shed light on citizen expectations and reservations 
regarding the evolution of their living environment \cite{reddel2004consultation}.

Finally, the \textit{scientific literature} broadens the search through 
conceptual and applied research on server deployment. Although not always 
realistic, these contributions support innovation by articulating strategies 
around clearly defined objectives, often intersecting with technological 
challenges \cite{gamatie2023}. Locations such as lampposts \cite{aral2021ares}  
and bus stops \cite{madamori2021enabling} 
are identified in this way.

Together, these four sources enable the construction of a coherent set of 
potential urban locations for distributed server deployment. Identification thus 
becomes an exploratory process in which candidate locations are progressively 
revealed through the analysis of diverse documentary sources.

The typologies identified must, as noted above, be compatible with the territory 
of interest. Territorial adaptation is therefore possible and encouraged: if 
metro station rooftops are identified as potential locations 
\cite{madeinmarseille_rtm_rails_pv}, a city served only by tramways may 
consider tram stations instead. Such adaptation is essential to avoid 
overlooking relevant site types: the broader the glossary, the greater the 
flexibility of deployment, which in turn supports a 
resilience-oriented 

approach toward urban limits \cite{chaoui2025integrating}. All 
typologies identified through this process share the characteristics of being 
anthropised surfaces, largely non-conflictual, compatible with local 
photovoltaic supply, and integrable without significantly extending the digital 
material footprint.

\paragraph{\bf Selection}

Not all identified locations automatically qualify for the glossary. A 
selection is required to discard typologies that are structurally incompatible 
with server deployment, based on four criteria: material compatibility, 
structural regulations, site specification, and site knowledge.

The \textit{material} criterion concerns incompatibilities between the server 
morphology and the physical characteristics of the site. These may arise from 
site types that compromise server operation, insufficient structural capacity 
to support a server, or inadequate space to accommodate computing units.

The \textit{structural} criterion reflects the strict regulations governing 
urban development \cite{talen2012city}. A deployment may conflict with these 
rules: in France, for instance, listed historical buildings are protected and 
require authorisation from the \textit{Architectes des Bâtiments de France} 
for any modification to their exterior appearance \cite{codepatrimoine_L6322}, 
making the installation of digital infrastructure on such buildings particularly 
difficult.

The \textit{specification} criterion addresses the risk of over-broad 
categories. Some location types are too heterogeneous to be expressed 
uniformly: scientific publications may propose deployments at network access 
points \cite{loven2020scaling, lahderanta2021edge}, which can take diverse 
forms — antennas, routers, and so on — making it impossible to assess 
material compatibility across the category as a whole.

Finally, the \textit{knowledge} criterion concerns data availability. 
Selecting appropriate locations requires a sufficiently comprehensive view of 
the territory \cite{castelnovo2016smart}; without reliable data describing the 
geographic position of sites, it is only possible to work on a case-by-case 
basis, precluding any territorial-scale assessment.

\subsection{Characterisation of identified locations}

The second step serves as a decision-support tool for identifying, within the 
glossary, the most favourable locations for server deployment. Rather than 
optimising classical computational metrics — latency, load balancing, and so on 
— it aims to determine which locations: (1) can meet the defined digital demand; 
(2) provide a local response to a local need, thereby remaining within context 
without drawing on external resources; and (3) enable server deployment while 
leveraging the advantages of each typology and acknowledging its constraints. 
Three criteria structure this characterisation: energy, spatial, and 
qualitative.

\paragraph{\bf Energy criterion}

The energy criterion assesses each location typology's capacity to generate 
sufficient photovoltaic energy to sustain a digital service, in keeping with 
the objective of minimising environmental impact. A photovoltaic (PV) panel 
converts shortwave
solar irradiance into electricity through semiconductor cells: when photons strike these cells, they excite electrons and generate an electric 
current \cite{bagnall2008photovoltaic}. Panel efficiency depends primarily on 
cell quality, which governs the conversion ratio between received irradiance 
and produced electricity — typically ranging from 12\% to 17.5\% depending on 
the technology selected \cite{parthiban2022enhancement} — and on the solar 
irradiance received, which is itself affected by geographic position, cloud 
cover, and shading \cite{rathod2016analysis}.

Geographic position determines the angle between the sun and the receiver as 
well as the distance separating them, both of which condition the received 
irradiance \cite{ahmad2011solar}. Combined with established models of the sun's 
apparent trajectory \cite{lefevre2013mcclear, ineichen2002new, 
gueymard1989two}, 

the location of the studied sites allows for estimating the cumulative irradiance over a given period of time (i.e., the irradiation named irradiation\textsubscript{cs}) under clear-sky conditions. 

Applying a conversion rate of 15\% yields a theoretical energy production estimate under optimal conditions (see Eq.~\ref{eq:eq1}):
\begin{equation}
    \text{prod} = 0.15 \times \text{irradiation}_\text{cs}
    \label{eq:eq1}
\end{equation}

Under real conditions, cloud cover reduces the fraction of irradiance reaching 
PV panels. We therefore introduce in Eq.~\ref{eq:eq2} a variable $neb \in [0, 1]$ representing this fraction:
\begin{equation}
    \text{prod} = 0.15 \times \text{irradiation}_\text{cs} \times \text{neb}
    \label{eq:eq2}
\end{equation}

Urban environments are further characterised by heterogeneous building heights 
and forms that generate variable shading conditions. To account for this, we 
introduce in Eq.~\ref{eq:eq3} a binary shading variable $sh$, set to $0$ if the site is shaded during the considered hourly period and $1$ otherwise:
\begin{equation}
    \text{prod} = 0.15 \times \text{irradiation}_\text{cs} \times \text{neb} \times \text{sh}
    \label{eq:eq3}
\end{equation}

Accounting for these three factors enables the estimation of PV energy 
production for each location typology. Given a digital demand and its 
associated energy consumption, it then becomes possible to right-size a 
distributed server network across the territory, creating a response 
environment for digital needs that minimises the environmental impact of 
computing energy consumption.

\paragraph{\bf Spatial criterion}

The spatial distribution of potential locations conditions how the 
infrastructure articulates with the city. We define a coherent spatial 
distribution through the principle of locality between digital demand and the 
infrastructure meeting it — expressed as the capacity to meet one's needs 
without infringing on the capacity of one's neighbours to meet theirs 
\cite{brundtland1987our}. In practice, this translates into analysing the 
distance between proposed locations and digital demand: by minimising these 
distances, we seek to foster spatial equity \cite{tan2017effects}, limit 
energy losses associated with data and energy transport \cite{bialek1996tracing}, 
and reduce latency. This framing also opens a discussion on forms of digital 
sufficiency that may emerge when needs cannot be satisfied locally 
\cite{jungell2022sufficiency}.

Distance analysis is operationalised through two complementary indicators. The 
first estimates the maximum distance required for one or more location 
typologies to meet a given digital demand, enabling comparison across typologies 
and prioritisation of those requiring the smallest travel distances. The second 
examines how the capacity to satisfy demand varies as a function of distance, 
revealing the irregular spatial availability of locations across the territory. 
A typology may require a large maximum distance to meet total demand while 
still satisfying a substantial fraction of it at short range — a finding that 
can support a redefinition of the initial digital demand toward lower levels 
that can be met as locally as possible, thereby anchoring digital sobriety 
within a spatial reflection on urban limits.

\paragraph{\bf Qualitative criterion}

The urban territory cannot be reduced to quantitative criteria alone 
\cite{forrester1970urban}. Our methodology therefore incorporates qualitative 
aspects that capture the intrinsic properties of location typologies in the 
context of server deployment, along two axes of common qualification and a set 
of typology-specific properties.

The first axis is \textit{environmental}, focusing on the structural 
modifications required to deploy distributed servers: the works needed to 
render a site compatible and the need to connect existing buildings or 
infrastructure to communication and electrical networks. In the absence of 
such connections, additional environmental and economic costs are incurred 
\cite{stidcurrent}. Connection to the main electrical grid also offers a 
twofold advantage: it enables the valorisation of surplus generation through 
resale and ensures continuity of supply during periods of solar intermittency.

The second axis concerns \textit{deployment acceptability}, captured through 
the number and nature of owners involved \cite{hosford2024acceptability}. 
Engaging multiple stakeholders is inherently more complex than dealing with a 
single one, whether at the scale of a single co-owned building requiring 
unanimous resident approval or across multiple locations divided between public 
and private actors. Furthermore, the conditions differ substantially depending 
on the nature of ownership: a French municipality may perceive collective or 
environmental value in hosting computing units, whereas a private individual 
may be deterred by costs, lack of expertise, or concern over impacts on their 
property.

These two axes yield six criteria across which typologies are scored. Assuming 
equal importance across criteria, each (criterion, typology) pair receives a 
score from the discrete set $\{0;\, 0.5;\, 1\}$, where $1$ indicates strong 
performance, $0.5$ average performance, and $0$ weak performance. This scoring 
permits a hierarchical ranking of typologies.

Beyond these common properties, each typology may also present specificities 
that are crucial to consider but do not permit fair comparison across 
categories. This limitation stems from the inherently unique nature of urban 
situations: a fully systemic treatment of the urban territory is not possible, 
and a situated consideration of the territory and its actors remains necessary 
to assess the relevance of the locations selected.
\section{Case study}
\label{sec:case-study}

In the next, we first describe the experimental setup of the case study. Then, we show the typical results provided by our methodology. 

\subsection{Experimental setup}

We illustrate our methodology through a case study conducted in Montpellier 
(INSEE code~34172), a city in southern France covering 56.88~km$^2$ with a 
population of 516{,}657 inhabitants.

\paragraph{\bf Digital demand.}
The digital demand studied here is the use of large language models (LLMs) in secondary education. Chatbots have been recently adopted by high-school students for learning, writing, information retrieval, and similar academic tasks \cite{stohr2024perceptions}. Unstructured and unrestricted use of these technologies can be harmful: it has been linked to lower academic performance, reduced neural connectivity patterns, and memory erosion. When their use is framed by a strict pedagogical protocol, the opposite effects are observed, with gains in learning and motivation \cite{kestin2024ai, krupp2023unreflected, forero2023chatgpt}. Three conditions make this difference: the model is used as a guide rather than a source of direct answers, it is consulted only after an initial phase of independent thinking, and it is restricted to low-level support such as spell-checking. Under close supervision by teaching staff, classroom use of chatbots can therefore deliver benefits.

Deploying an LLM by and for high schools is well suited to this framing. LLM inference is, however, energy-intensive: the electricity consumption of the infrastructure needed to meet AI demand is projected to grow by a factor of two to fourteen between 2025 and 2035 \cite{paccou:hal-05015754}. At that scale, the infrastructure supporting conversational agents risks conflicting with current environmental goals such as limiting global warming.

LLMs can thus play an active role in school education, but only under conditions. Their deployment cannot be planned on the assumption of unlimited resources. Reconciling the social benefits of these tools with global constraints requires an infrastructure that supports their use within environmental limits, and a usage regime designed to limit rebound effects.

\begin{figure}[htbp]
    \centering
    \includegraphics[width=0.38\textwidth, trim={0 2cm 0 1.5cm},clip]
        {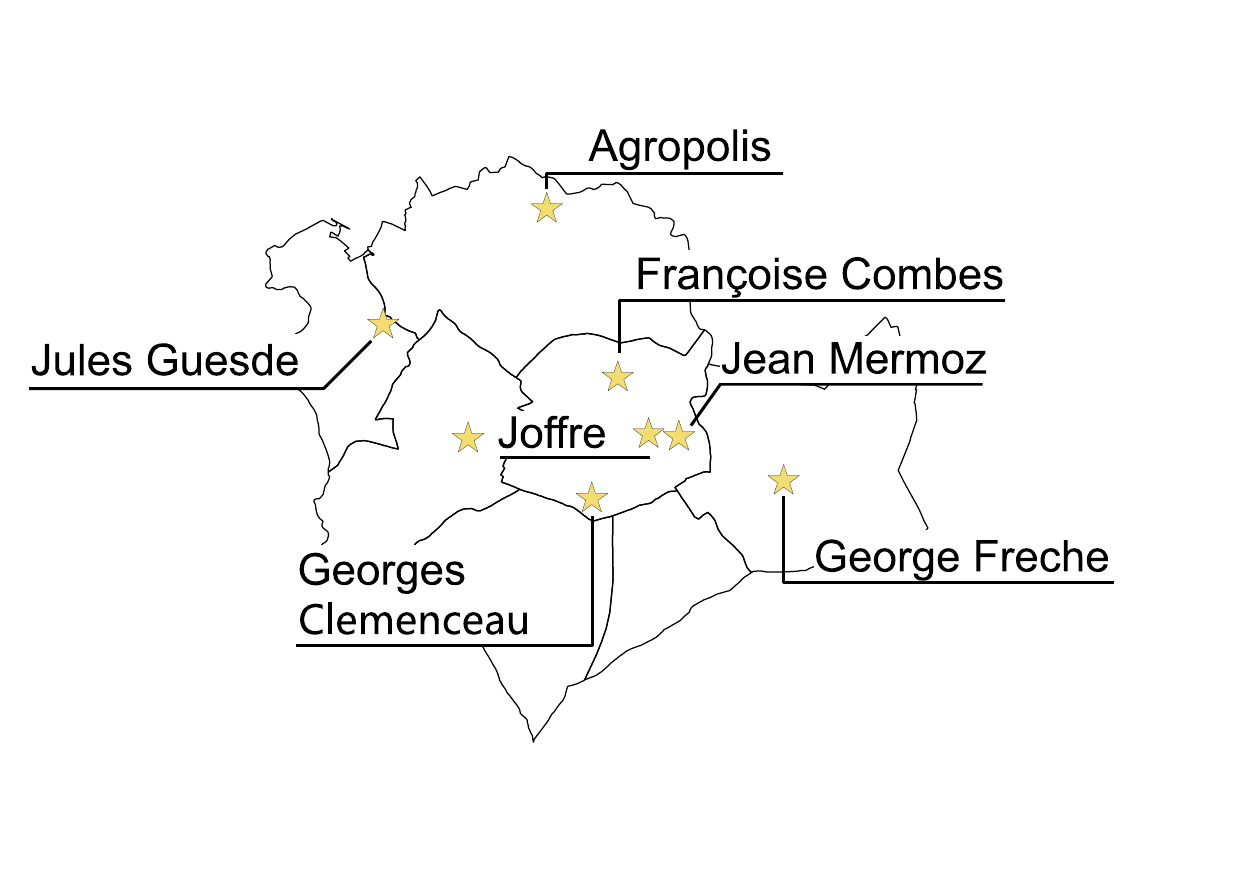}
    \caption{City of Montpellier and its public high schools.}
    \label{fig:montpellier_lycee}
\end{figure}

The demand is centered on the seven public high schools of Montpellier 
(Figure~\ref{fig:montpellier_lycee}): Joffre, George~Clemenceau, Jules~Guesde, 
Jean~Mermoz, Françoise~Combes, Agropolis, and George~Frêche. We define a 
scenario in which 10\% of students use an LLM between 08:00 and 18:00 (UTC+1) on 
21~December (winter solstice). According to the French Ministry of 
Education, this corresponds to 158, 115.2, 197.2, 177.4, 25.2, 39.1, and 
12.8 active users per school, respectively.

\paragraph{\bf Computing infrastructure.}
We adopt a computing unit morphology based on the Genesis 
paradigm~\cite{gamatie2023}. This approach allows digital infrastructure to 
take various forms so as to adapt to different location typologies while being 
equipped with a solar panel and able to operate outdoors. To quantify the 
energy required to meet the digital demand of the high schools, we use a 
Genesis server prototype deployed in Montpellier, equipped with an Intel Xeon 
E3 3.3\,GHz processor and 16\,GB of RAM, without a GPU or TPU. Based on a consumption study, we determine an average power draw of 502\,Wh when running the 
\href{https://huggingface.co/meta-llama/Llama-3.1-8B-Instruct}{Llama~3.1~8B} 
model. All processing is carried out in a Python~3.10.12 environment.

We chose not to include an energy-storage system in the target system design, so that servers operate only during daylight hours. Two reasons motivate this choice. 
First, urban energy storage relies mainly on batteries \cite{anastasovski2024energy}. The extraction of the required resources, the manufacturing of the devices, and their transport carry heavy environmental costs -- in particular for the movement of raw materials \cite{mcmanus2012environmental} -- and heavy social costs, including the forced displacement of populations linked to certain mining activities \cite{mandishekwa2020mining}. 
Second, restricting access to daylight hours regulates usage and reduces the risk of addiction. Continuous energy availability would encourage intensive LLM use during periods normally reserved for student rest, which is necessary for the balance of daily rhythms \cite{yeo2020associations}.

\paragraph{\bf Location typologies.}
We identify candidate hosting locations through French legislation and 
regulation~\cite{code_environnement, code_energie, code_urbanisme, 
code_construction_habitation}, urban projects deploying photovoltaic panels on 
existing infrastructures~\cite{route_solaire_wikipedia, colas_wattway_pays_bas, 
made_in_marseille_metro_rose, zgsm_solar_traffic_lights_portugal, 
pv_magazine_mur_antibruit, fonroche_senegal_50000, spie_batignolles_coasis, 
urbaneo_helios_solaire, ciel_terre_leutenheim_flotta, jpee_inauguration}, and 
scientific studies  \cite{lahderanta2021edge, laha2020, aral2021, madamori2021enabling, gamatie2023}. 
Citizen consultation documents are omitted 
owing to limited data availability. After search and selection, sixteen 
location typologies are retained: lamppost, individual housing, collective 
housing, service building, bench, parking area, industrial building, bus 
station, traffic light, table, network antenna, bicycle facility, sports 
building, tram station, brownfield, and polluted land.

For each typology, all instances within the administrative boundaries of 
Montpellier are modeled in Python with \textit{Geopandas} \cite{kelsey_jordahl_2020_3946761}, including their geographic position and, where applicable, their geometric footprint (Table~\ref{table:glossaire}). For typologies lacking geometric data, contextually coherent installation areas are 
assigned (Table~\ref{table:surface_instalation}).

\begin{table}[t]
    \centering
    \caption{Glossary of identified and modeled location typologies.}
    \small
    \begin{tabular}{|l|r|l|}
        \hline
        Location type         & Count    & Data source \\ \hline\hline
        Base station          & 599      & \href{https://data.arcep.fr/mobile/sites/last/}{\textit{Mon Réseau Mobile}} \footnotemark[1] \\ \hline
        Bench                 & 4{,}259  & \href{https://www.openstreetmap.org/}{OSM} \footnotemark[2] \\ \hline
        Industrial building   & 1{,}207  & \href{https://geoservices.ign.fr/bdtopo}{BD~TOPO®} \footnotemark[3] \\ \hline
        Service building      & 12{,}728 & \href{https://geoservices.ign.fr/bdtopo}{BD~TOPO®} \\ \hline
        Sports building       & 251      & \href{https://geoservices.ign.fr/bdtopo}{BD~TOPO®} \\ \hline
        Traffic light         & 1{,}638  & \href{https://www.openstreetmap.org/}{OSM} \\ \hline
        Industrial wasteland  & 100      & \href{https://www.data.gouv.fr/datasets/sites-references-dans-cartofriches}{\textit{CartoFriche}} \footnotemark[4] \\ \hline
        Streetlight           & 87{,}449 & \href{https://data.montpellier3m.fr/}{\textit{Open Data Montpellier}} \footnotemark[5] \\ \hline
        Bike storage room     & 264      & \href{https://www.openstreetmap.org/}{OSM} \\ \hline
        Multi-family housing  & 18{,}645 & \href{https://geoservices.ign.fr/bdtopo}{BD~TOPO®} \\ \hline
        Single-family housing & 48{,}037 & \href{https://geoservices.ign.fr/bdtopo}{BD~TOPO®} \\ \hline
        Parking area          & 2{,}126  & \href{https://www.openstreetmap.org/}{OSM} \\ \hline
        Polluted site         & 42       & \href{https://www.georisques.gouv.fr/donnees/bases-de-donnees/sites-et-sols-pollues-ou-potentiellement-pollues}{\textit{Sites et sols pollués}} \footnotemark[6] \\ \hline
        Bus stop              & 816      & \href{https://www.openstreetmap.org/}{OSM} \\ \hline
        Tram stop             & 297      & \href{https://www.openstreetmap.org/}{OSM} \\ \hline
        Table                 & 712      & \href{https://www.openstreetmap.org/}{OSM} \\ \hline
    \end{tabular}
    \label{table:glossaire}
\end{table}

\footnotetext[1]{\textit{Mon Réseau Mobile} details the list of mobile network sites in France.}
\footnotetext[2]{OpenStreetMap (OSM) data consists of collaboratively contributed geospatial features describing a wide range of real‑world geographic entities.}
\footnotetext[3]{The BD TOPO® dataset is a database describing France’s infrastructure.}
\footnotetext[4]{\textit{CartoFriche} lists potentially unused or derelict artificialized sites across France.}
\footnotetext[5]{\textit{Open Data Montpellier} provides open access to a wide range of public datasets from the city}
\footnotetext[6]{\textit{Sites et sols pollués} identifies polluted or potentially polluted sites in France.}

\begin{table}[H]
    \centering
    \caption{Proposed installation area for location typologies without 
    geometric data.}
    \small
    \begin{tabular}{|l|r|}
        \hline
        Location typology & Installation area (m\textsuperscript{2}) \\ \hline\hline
        Base station      & 3  \\ \hline
        Bench             & 2  \\ \hline
        Table             & 2  \\ \hline
        Streetlight       & 1  \\ \hline
        Bus stop          & 4  \\ \hline
        Tram stop         & 4  \\ \hline
        Bike storage room & 4  \\ \hline
    \end{tabular}
    \label{table:surface_instalation}
\end{table}

\paragraph{\bf Solar irradiance and shading.}
Clear-sky irradiance is estimated at hourly intervals using the PVLIB 
library~\cite{jensen2023pvlib} and the Ineichen model~\cite{ineichen2002new}.
Shading is computed from a digital surface model provided by the French 
national geographic institute, following the methodology of Corripio~\cite{corripio2003vectorial} with insolation a Python library \cite{insolation}; each location is assigned a binary shading variable (0 if shaded, 1 otherwise) at each hour. To account for cloud 
cover, data from \textit{Météo France} describing the number of sunny hours per 
month are extrapolated to an intra-day distribution, concentrating cloud-free 
periods at peak irradiance hours (11:00--15:00 UTC+1).

\subsection{Results}

\begin{figure}[htbp]
    \centering
    \includegraphics[width=0.42\textwidth]{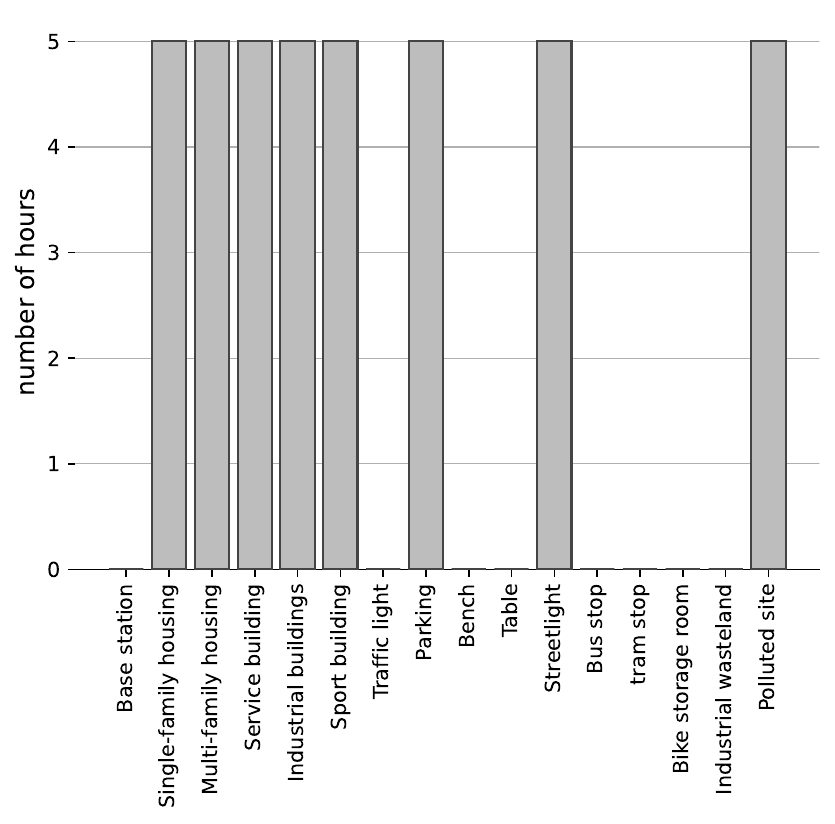}
    \caption{Number of hours for which each location typology can sustain the 
    LLM energy demand.}
    \label{fig:soutenir_energie}
\end{figure}

\paragraph{\bf Energy feasibility.}
Under the meteorological scenario described above, eight location typologies 
are able to sustain the energy demand associated with the LLM 
(Figure~\ref{fig:soutenir_energie}). Each of these typologies produces 
sufficient energy during the five hours of peak sunlight.

\paragraph{\bf Proximity to demand.}
Among the energy-feasible typologies, we assess the ability of each to meet 
the digital demand locally. A linear solver minimizes the distance between 
candidate locations and high schools while ensuring the energy required for 
operating the computing units is satisfied without exceeding the site’s capacity.

\begin{table*}[t]
    \centering
    \caption{Qualitative analysis of selected location typologies.}
    \small
    \begin{tabular}{|l|c|c|c|c|c|}
        \hline
        & Network & Energy & Structure
        & \begin{tabular}[c]{@{}c@{}}Number of owners\end{tabular}
        & \begin{tabular}[c]{@{}c@{}}Nature of  ownership\end{tabular} \\ \hline\hline
        Service building      & 1 & 1 & 1 & 1   & 1   \\ \hline
        Single-family housing & 1 & 1 & 1 & 0   & 0   \\ \hline
        Multi-family housing  & 1 & 1 & 1 & 0   & 0.5 \\ \hline
        Parking area          & 1 & 1 & 1 & 0.5 & 0.5 \\ \hline
    \end{tabular}
    \label{table:analyse_qualitative}
\end{table*}

\begin{figure}[htbp]
    \centering
    \includegraphics[width=0.45\textwidth]{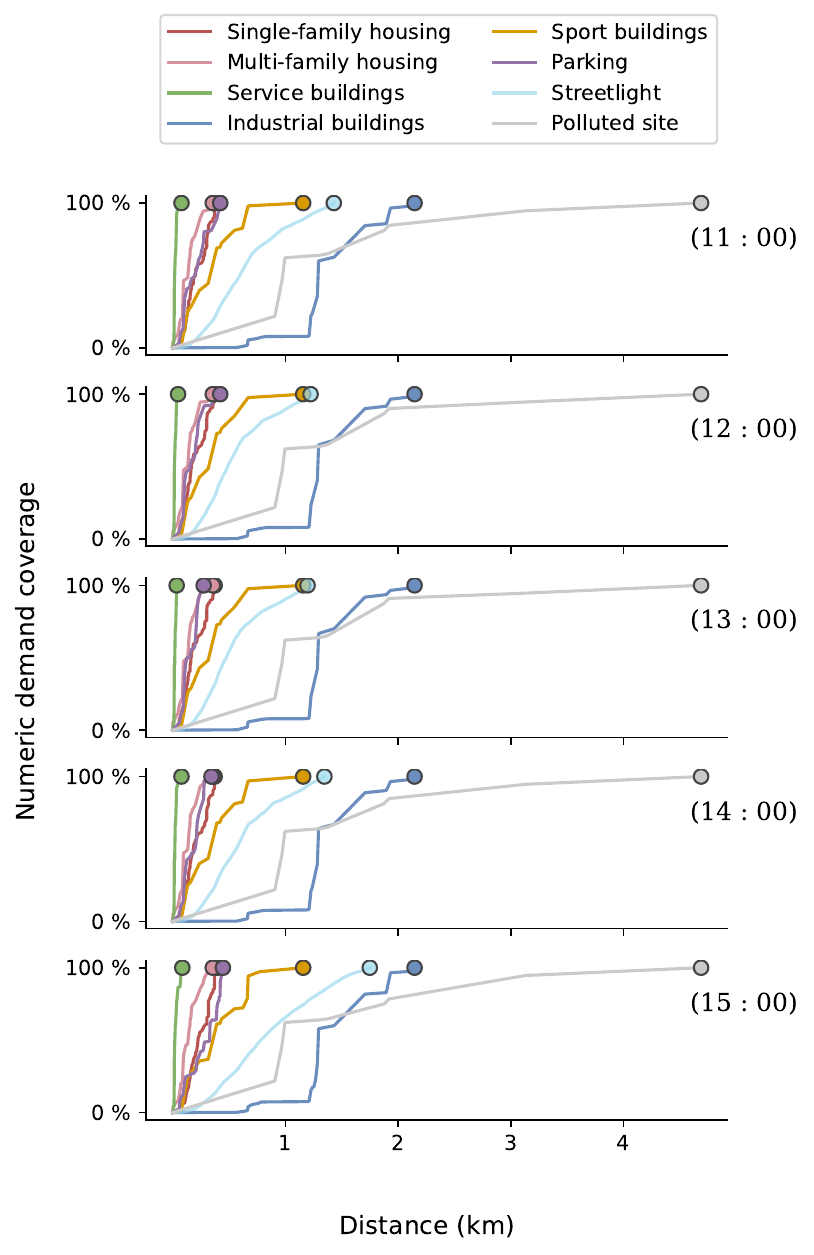}
    \caption{Cumulative digital demand coverage as a function of the maximum 
    distance to demand, shown for each energy-feasible location typology and 
    for each peak solar hour (11:00--15:00 UTC+1). The filled circle on each curve 
    marks the minimum distance at which that typology achieves full demand 
    coverage.}
    \label{fig:distance_demande}
\end{figure}

Figure~\ref{fig:distance_demande} presents, for each of the five peak solar 
hours (11:00--15:00 UTC+1), the cumulative digital demand coverage as a function of 
the maximum distance allowed between a computing unit and a high school. Each 
curve corresponds to one location typology; the filled circle at its right 
extremity marks the minimum distance at which that typology alone achieves 
full demand coverage. Reading a curve from left to right therefore reveals how 
quickly — in spatial terms — a typology can respond to the totality of the 
digital demand: a steep rise close to the origin indicates that a large share 
of demand can be met at very short range, while a gradual slope or a late 
saturation point signals a typology whose suitable instances are sparsely 
distributed relative to the demand sites. The replication of this 
representation across five hourly panels reflects the temporal variability of 
photovoltaic production: the number of computing units that can actually be 
powered at a given location changes as solar irradiance evolves throughout the 
day, which in turn affects how much demand each typology can cover at a given 
distance. The overall spatial ranking among typologies nevertheless remains 
stable across all five hours, attesting to the robustness of the conclusions 
drawn below.

Two groups of typologies emerge clearly. The first comprises service buildings, 
single-family housing, multi-family housing, and parking areas, all of which 
reach full coverage within less than 500~m across every hourly panel. Service 
buildings saturate at the shortest distances — between 35 and 85~m depending 
on the hour — because the rooftops of the high schools themselves belong to 
this typology, making the spatial gap between demand and supply negligible. 
The remaining three typologies in this group achieve full coverage between 274 
and 445~m, indicating that sufficiently dense instances of each exist in the 
immediate urban fabric surrounding the demand sites. The second group — 
industrial buildings, sport buildings, streetlights, and polluted sites — 
requires distances exceeding 1~km to satisfy total demand, with polluted sites 
reaching saturation only beyond 4.5~km. Given the availability of closer 
alternatives, these typologies are excluded in order to preserve the local 
response strategy central to our methodology.

\paragraph{\bf Qualitative assessment.}
The four selected typologies are then evaluated qualitatively through a 
hierarchical scoring based on the need to connect locations to power and 
internet networks, required structural modifications, number of distinct 
owners, and nature of ownership (Table~\ref{table:analyse_qualitative}).

This evaluation places service buildings in first position. In the specific 
context of public high schools, deploying computing units on municipal service 
buildings is particularly appropriate: it avoids reliance on external private 
actors and eliminates the risks associated with contractual interruptions or 
rental costs incompatible with educational budgets. Accordingly, this case 
study recommends deployment on service buildings, and more specifically on 
those owned by the municipality.

Each region and each stakeholder has specific characteristics that the model cannot fully capture. The proposed model is an approximate representation built on a proximity logic; it does not claim to be a universal solution but a starting point for discussion with the potential stakeholders. They hold the detailed knowledge of needs and constraints that no modeled framework can generalize, and dialogue with them is needed to adjust the locations to be deployed.

\section{Discussion on limitation of this work}
Our methodology has a scope and a set of conditions that should be made explicit. It follows the spirit of LIMITS by treating digital infrastructure not as an abstract optimisation problem but as a socio-technical object embedded in finite urban, material, and ecological conditions. It works within urban limits by relying only on existing infrastructures and already anthropized surfaces, by integrating structural, social, and energy constraints, and by seeking proportionality between digital demand and the computing capacity actually deployed.

A first limitation is the strong dependence on available data. In France, most relevant data come from government sources, which generally ensures accessibility and reliability. In other contexts, such data may be restricted or absent. International open data sources such as OpenStreetMap partly compensate, but their quality and completeness depend on user contributions and remain uneven. The methodology also depends on the territorial context in which it is applied: data availability, regulatory frameworks, urban morphology, ownership structures, and local energy resources all shape its transferability. It should therefore be read as a situated approach, to be locally adapted rather than directly replicated.

A second limitation concerns the spatial assumptions of the model, which presupposes that rooftops and other anthropized surfaces can be mobilized. Such spaces may already host other technical devices -- solar panels, ventilation systems, or equipment needed for the building's normal operation -- or support uses that contribute to urban sustainability, such as greening, urban agriculture, or shared gardens. As the paper notes, deploying computing capacity on existing infrastructures may conflict with the primary functions of places that already serve social, residential, or collective purposes.

A third limitation is that our analysis considers the pedagogical benefits of LLM use without examining the full chain of involved  actors. Teachers are a first essential group: integration requires pedagogical appropriation, adaptation of practices, and reflection on the practical conditions of classroom use. Citizens and residents are directly concerned as well, since the installation of computing units may create local nuisances such as noise, heat, or conflicts of use. Public authorities and urban stakeholders face economic constraints linked to the financing, deployment, maintenance, and renewal of these infrastructures, which is a major obstacle to implementation. As the paper also recalls, solar intermittency increases maintenance needs and limits continuous operation.

Our study is limited to a single digital demand of a pedagogical nature, defined through a specific scenario. This framing makes the methodology testable but restricts the scope of the results. Extending the approach to other urban needs and other actors — housing providers, municipalities, citizens, public services, associations, technical operators — would require treating these infrastructures as supporting a broader community of uses and energy flows \cite{soden2022photovoltaic}. Such an extension would change the management and governance questions: the challenge would no longer be only to identify a technically feasible infrastructure, but also to determine how access, contribution, maintenance, and decision-making are distributed among actors with potentially divergent interests. Territorial equity, socio-spatial justice, and collective governance would then become central to the long-term durability of this form of local digital sovereignty. The paper already integrates some dimensions of acceptability and equity but does not yet fully address these governance issues.

\section{Conclusion and perspectives}
\label{sec:conclu}

Our work highlights a central insight regarding the deployment of digital technologies within the urban territory: a frugal digital service is not merely a matter of efficiency, but a matter of situated adequacy between needs, existing infrastructures, and local resources. By relying exclusively on already developed locations and on local photovoltaic production, we show that it is possible to support a digital service while respecting the environmental and social limits of the territory.

The approach we adopted sheds light on the types of urban infrastructures that are most relevant for sustaining a distributed energy demand, both in terms of spatial availability, energy production capacity, and compliance with the social and structural constraints of the territory. Among the sixteen typologies initially identified, only a subset displayed sufficient capacity to contribute meaningfully to powering servers. Their use nevertheless depends on practical, institutional, and territorial considerations, especially when the objective is to maintain a local deployment strategy.

The intrinsic robustness of the proposed framework stems from three complementary factors. First, the multiplicity of infrastructures—ranging from rooftops to streetlights—ensures wide spatial coverage and redundancy. Second, adherence to territorial constraints, including structural, social, and local digital demand considerations, anchors the deployment in realistic and acceptable urban contexts. Third, leveraging solar energy as a local resource reduces dependence on external energy supplies. Together, these elements prioritize robustness over peak performance, offering a sustainable digital service capable of withstanding a highly uncertain world marked by energy disruptions, conflicts, and other emerging risks.

Beyond the experimental scope, this work highlights the importance of systematically integrating energy, spatial, and social criteria into the planning of emerging digital infrastructures. It also shows that distributed models supported by local photovoltaic resources and by the reuse of existing buildings constitute a promising avenue for reconciling digital usage with the reduction of associated environmental impacts. Such an approach opens the way to deployment strategies that are more robust, more sustainable, and more closely aligned with urban dynamics.

In light of these results, several avenues for future research emerge. Applying this strategy to traditional studies on the optimization of computing‑unit deployment (see Section \ref{sec:related}) would ground the analysis in a realistic urban context while clarifying the types of digital services whose quality‑of‑service requirements could be met under such conditions. 

To avoid rebound effects, the evaluation metrics used in the literature also need to be reconsidered, so that they reward properties associated with sustainability rather than optimization alone. To avoid presenting any single solution as normative, results should be reported in a nuanced form -- for example as a Pareto front \cite{deb2011multi} -- that makes visible the trade-offs of prioritizing one sustainability metric over another.

Another research avenue concerns the methodology’s ability to withstand urban change. Applying it to the same area at two different points in time would make it possible to evaluate the long‑term relevance of an initial deployment.

Finally, as mentioned earlier in the paper,  future research directions should rely on active dialogue with urban stakeholders to identify desirable digital uses together with their needs and constraints, and to quantify them. Exploratory interviews \cite{blanchet2007enquete} are one suitable instrument and allow different types of digital uses to be distinguished at an early stage. The relevance of the identified needs should then be assessed against a sustainability frame, drawing for instance on \cite{nardi2018computing}, which considers computing through planetary boundaries (natural resources, energy, climate), or on \cite{hamant2023antidote}, which favors systems able to absorb fluctuations over those optimized for efficiency. Because each region and each stakeholder has specific characteristics, the proposed model -- built on a proximity logic -- is only an approximate representation and is not intended as a universal solution; it is a starting point for discussion with the actors who hold the detailed knowledge of needs and constraints that modeled frameworks can hardly capture.

\section*{Acknowledgments}
This work has been funded by the CNRS-MITI under the Explainer project grant. 
The authors used Generative AI tool for language editing and grammar checking. All technical content and results were produced and verified by the authors. 

The authors would like to thank all reviewers for their valuable comments and suggestions, which helped us to improve the quality of the paper.

\bibliographystyle{ACM-Reference-Format}
\bibliography{biblio}

\end{document}